\documentclass[aps,prl,twocolumn,showcaps,superscriptaddress,showpacs,amsmath,amssymb]{revtex4}

\usepackage{graphicx}
\usepackage{dcolumn}
\usepackage{bm}
\usepackage[dvips]{epsfig}
\usepackage{color}
\newcommand{\tauCorr}{\ensuremath{\tau_\mathrm{corr}}\,}
\newcommand{\tauR}{\ensuremath{\tau_\mathrm{r}}\,}

\newcommand{\kT}{k_\mathrm{B}T}

\newcommand{\mean}[1]{\langle #1 \rangle}

\newcommand{\unit}[1]{\;\mathrm{#1}}
\newcommand{\RM}[1]{\MakeUppercase{\romannumeral #1}}
\newcommand{\NESSI}{NESS~\MakeUppercase{\romannumeral 1}\,\,}
\newcommand{\NESSII}{NESS~\MakeUppercase{\romannumeral 2}\,\,}
\newcommand{\psI}{\ensuremath{p\mathrm{^s_{\RM{1}}}}\,}
\newcommand{\psII}{\ensuremath{p\mathrm{^s_{\RM{2}}}}\,}

\newcommand{\vII}{\mean{v}\mathrm{_{\RM{2}}}}
\newcommand{\fI}{\ensuremath{f\mathrm{_{\RM{1}}}}\,}
\newcommand{\fII}{\ensuremath{f\mathrm{_{\RM{2}}}}\,}
\newcommand{\tI}{\ensuremath{t\mathrm{_{\RM{1}}}}\,}
\newcommand{\tII}{\ensuremath{t\mathrm{_{\RM{2}}}}\,}
\newcommand{\taueq}{\ensuremath{\tau\mathrm{_r^{eq}}}\,}

\begin{document}

\title{Relaxation of a Colloidal Particle into a Nonequilibrium Steady State}

\author{Valentin Blickle}
\affiliation{2. Physikalisches Institut, Universit\"at Stuttgart,
Pfaffenwaldring 57, 70550 Stuttgart, Germany}
\author{Jakob Mehl}
\affiliation{2. Physikalisches Institut, Universit\"at Stuttgart,
Pfaffenwaldring 57, 70550 Stuttgart, Germany}
\author{Clemens Bechinger}
\affiliation{2. Physikalisches Institut, Universit\"at Stuttgart,
Pfaffenwaldring 57, 70550 Stuttgart, Germany}
\affiliation{Max-Planck-Institut f\"{u}r Metallforschung,
Heisenbergstrasse 3 ,70569 Stuttgart, Germany}

\begin{abstract}
We study the relaxation of a single colloidal sphere which is
periodically driven between two nonequilibrium steady states.
Experimentally, this is achieved by driving the particle along a
toroidal trap imposed by scanned optical tweezers. We find that the
relaxation time after which the probability distributions have been
relaxed is identical to the decay of the velocity autocorrelation
function, measured in a steady state. In quantitative agreement with
theoretical calculations the relaxation time strongly increases when
driving the system further away from thermal equilibrium.

\end{abstract}


\maketitle


The understanding of thermodynamic processes at small length scales
is of central importance at the interface of physics, biology and
chemistry. Classical thermodynamics, as originally developed for
macroscopic systems with many internal degrees of freedom cannot be
applied to e.g. molecular machines, proteins or micro-mechanical
devices. This is because at microscopic scales, thermal fluctuations
must not be neglected and the familiar well-defined thermodynamical
quantities have to be replaced by corresponding distributions of
finite width ~\cite{bus05, rit08, sei08}. The situation is further
complicated when these systems are driven out of thermal equilibrium
as often encountered within their natural environment. The treatment
of fluctuations in such nonequilibrium situations is even more
difficult since it requires the full knowledge of the system's
dynamics. Despite considerable progress in deriving exact
relationships which are valid beyond thermal
equilibrium~\cite{jar97, lip02, sei05} a comprehensive theoretical
description of nonequilibrium is still lacking.

Among the huge manifold of nonequilibrium conditions, nonequilibrium
steady states (NESS) are certainly the most simple conceivable
situations, being characterized by a time-independent probability
distribution in the presence of a non-vanishing probability current.
Accordingly, NESS present ideal conditions for fundamental studies
and tests of nonequilibrium properties~\cite{oon98} on a microscopic
scale~\cite{hat01, spe06}.

In this Letter we experimentally investigate the relaxation behavior
of a single colloidal particle which is periodically driven between
two different nonequilibrium steady states \NESSI and \NESSII being
created by scanning optical tweezers. We find that the NESS
relaxation time as defined by the decay of the probability
distribution only depends on the final state but is independent of
the initial one. In addition, we show that this relaxation time is
identical to that obtained by the decay of the velocity
autocorrelation function in the steady state regime, i.e. after
relaxation has been completed. In agreement with theoretical
calculations, the relaxation time increases when driving the system
further away from thermal equilibrium.


\begin{figure}
  \includegraphics{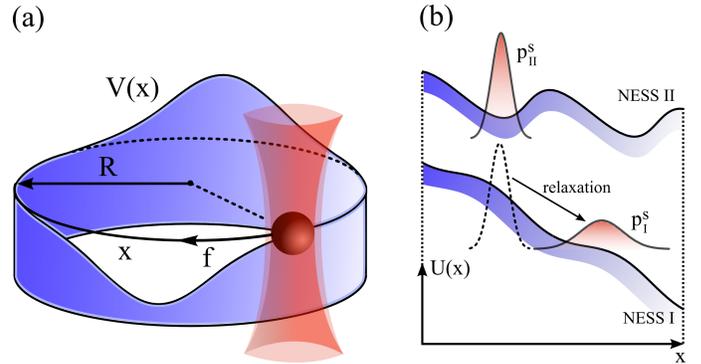}
\caption{(color online) (a) Realization principle of the creation of
a NESS for a colloidal particle by scanning a focussed laser beam.
(b) Schematic representation of \NESSI and \NESSII which correspond
to a tilted periodic potential. A sudden change of the driving force
$f$ and phase $\phi=\pi$ leads to a redistribution of the related
probability distributions \psI and \psII. \label{fig:system}}
\end{figure}

The experimental setup has been already described elsewhere and will
be discussed here only in brief ~\cite{bli07a}. Well-defined
nonequilibrium steady states for a colloidal silica particle
immersed in water, with radius $a=0.65 \unit{\mu m}$ are created by
scanning the highly focussed beam of a Nd:YAG laser ($\lambda =
532\unit{nm}$) along a circle with radius $R=1.14\unit{\mu m}$ (see
Fig.~\ref{fig:system}). At rather high scanning frequencies the
particle cannot follow the tweezers motion due to the viscous forces
of the fluid, and it is confined to an effective three-dimensional
toroidal optical trap. At intermediate scanning frequencies,
however, each time the scanning laser focus passes the particle, a
small displacement of the colloid along the scanning direction is
induced. Since individual kicks are not resolvable by digital video
microscopy \footnote{The particle trajectory is monitored with a
spatial and temporal resolution of $20\unit{nm}$ and $33\unit{ms}$,
respectively.}, in this regime the scanning tweezers can be
considered as exerting a constant force $f$ on the particle along
the angular coordinate $x$ ~\cite{fau95, lutz06, bli07a}. For a
scanning frequency of $200 \unit {Hz}$ and a laser intensity
$I_0\approx40 \unit{mW}$ this leads to a drift velocity of $v\approx
7 \unit {\mu m / s}$. In addition, the laser intensity is weakly
modulated along the toroidal trap with an electro-optical device
whose input signal is synchronized with the scanning motion of the
laser focus. For a periodic intensity modulation $I(x)=I_0+\Delta
I\sin(x)$ this leads to additional optical gradient forces, i.e. a
static potential $V(x)=-\frac{V_0}{2}\sin(x+\phi)$ acting on the
particle. The value of $\phi$ can be controlled by the relative
phase difference between the scanned tweezers motion and its
intensity variation. In total, the colloid is subjected to a tilted
periodic potential $U(x)=V(x)-fRx$ corresponding to a NESS where $f$
and $V(x)$ can be tuned by $I_0$ and $\Delta I$, respectively. The
driving force $f$ and $V(x)$ are not known {\it a priori} but can be
reconstructed via a generalized Boltzmann factor from the measured
stationary probability distribution $p^s(x)$ and the probability
current in the system ~\cite{bli07b}.

The relaxation of a colloidal particle into a NESS is investigated
by periodically toggling between two differing steady states. This
is accomplished by a sudden change in the driving force $f$ and the
phase $\phi$ according to the protocol
\begin{equation}
\begin{split}
&\text{if $0\leq t \leq \tI$:}\hspace{1.2cm} \fI,\,\,\phi_{\text{\RM{1}}}=0\hspace{0.2cm}\text{(\NESSI)}    \\
&\text{if $\tI< t\leq \tI+\tII$:}\hspace{0.2cm}
\fII,\,\phi_{\text{\RM{2}}}\hspace{0.7cm}\text{(\NESSII)}.
\end{split}
\label{eq:protocol}
\end{equation}

Unless otherwise stated, $V_0$ is kept constant at
$V_0\approx100\unit{\kT}$. The duration times $\tI$ and $\tII$ are
chosen sufficiently long to allow the system to reach the
corresponding stationary probability distributions \psI and \psII.
The entire protocol is typically repeated up to 800 times during
each experiment to obtain adequate statistical averages.


\begin{figure}
  \includegraphics{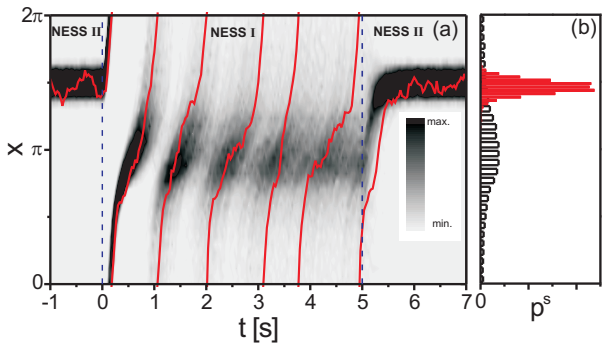}
\caption{(color online) (a) Particle trajectory (solid line) and
$p(x,t)$ as gray scaled background. Dashed vertical lines indicate
the transitions between two different NESS with durations
$\tI=5\unit{s}$, $\tII=2\unit{s}$ and a phase difference
$\phi_{\text{\RM{2}}}=\pi$. (b) Normalized steady state
probability distributions of \NESSI (open bars) and \NESSII
(closed bars).\label{fig:histo}}
\end{figure}


To illustrate the principle of our experiments, we first discuss the
situation where \NESSII is close to thermal equilibrium. This is
achieved by applying a rather weak driving force $\fII\approx
4\unit{\kT/\mu m}$. Accordingly, $U(x)$ exhibits a potential well of
about $80\unit{\kT}$ where the particle remains strongly localized,
thus closely resembling equilibrium conditions (locked state). This
is clearly seen by the trajectory (solid line) in
Fig.~\ref{fig:histo}(a) which is confined to a small range of
$x$-values. In contrast, \NESSI has a much stronger force $\fI
\approx 53\unit{\kT /\mu m }$. Therefore, $U(x)$ exhibits no local
minimum and the particle is free to drift along the entire torus
(running state).

Because of the superimposed Brownian motion, the particle trajectory
varies between each cycle of the protocol. These fluctuations are
taken into account by considering the probability distribution
$p(x,t)$, i.e. the probability of finding the particle at time $t$
at position $x$. The measured $p(x,t)$ obtained from about $800$
cycles of the protocol is shown as gray scaled background of
Fig.~\ref{fig:histo}(a). For $-1<t<0\unit{s}$ the particle has
relaxed to \NESSII where it is localized inside the deep potential
minimum. The corresponding strongly peaked steady state probability
distribution \psII is shown as closed bars in
Fig.~\ref{fig:histo}(b). Upon suddenly switching to \NESSI at
$t=0\unit{s}$, the particle starts to circulate along the entire
toroidal trap; this leads to a broadening and a shift of the maximum
in $p(x,t)$. The damped oscillatory behavior of $p(x={\rm
const.},t)$ is typical for the relaxation into a nonequilibrium
steady state. This is in contrast to the situation at $t=5\unit{s}$
when the protocol switches back to the equilibrium-like conditions
of \NESSII. Here, $p(x={\rm const.},t)$ monotonically approaches its
final value \psII. It should be noted that the relaxation from
\NESSI into \NESSII proceeds much more rapidly than into the other
direction.

\begin{figure}
  \includegraphics{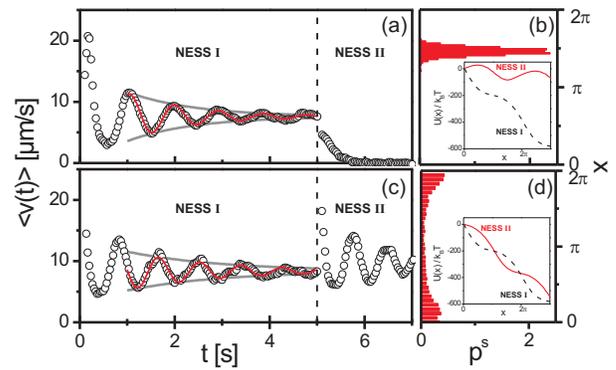}
\caption{(color online) (a),(c) Open symbols: Mean drift velocity
$\mean{v(t)}$ after switching from \NESSII to \NESSI and vice versa
($\fI \approx 53\unit{\kT /\mu m }$, $\phi_{\text{\RM{1}}}=0$). The
\NESSII parameters are given in Tab.~\ref{tab:zerfallszeit} (a):
(i), (c): (iv). Solid line: Exponentially damped sinusoidal
function. (b),(d): Corresponding potentials (inset) and steady state
distributions of \NESSII.\label{fig:zerfall}}
\end{figure}


To quantify our findings we calculate the mean drift velocity
$\mean{v(t)}=\mean{\frac{x(t+\Delta t)-x(t-\Delta t)}{2\Delta t}}$,
which is obtained by averaging the actual particle velocity $v(t)$
over several hundred cycles of the protocol. Since the length scale
over which the potential $U(x)$ varies is more than one order of
magnitude larger than the maximal particle displacement between two
consecutive ($\Delta t =33 \unit{ms}$) video frames, $\mean{v(t)}$
can be obtained from the experimentally determined trajectories. The
symbols in Fig.~\ref{fig:zerfall}(a) show $\mean{v(t)}$ for the same
data set as in Fig.~\ref{fig:histo}. After switching to \NESSI,
$\mean{v(t)}$ is a decaying oscillatory function which converges to
the corresponding mean steady state velocity. For $t>1\unit{s}$ it
can be well described by an exponentially damped sinusoidal function
(solid line) with decay time $\tau=1.4\pm 0.2\unit{s}$ and the
oscillation period given by the mean particle revolution time
$T_R=0.9\unit{s}$. Similar as above, $\mean{v(t)}$ for the
relaxation into the equilibrium-like \NESSII is purely exponential
with a decay time of $0.3\unit{s}$.

In order to understand how the relaxation into a NESS compares
with that into thermal equilibrium we consider the relaxation time
\taueq of an overdamped Brownian particle into a parabolic
potential. According to Ornstein-Uhlenbeck ~\cite{uhl30, gar04}
this is given by
\begin{equation}
\taueq=\frac{6\pi\eta a}{k}\label{eq:tauRgg}
\end{equation}
where $\eta$ is the viscosity of the solvent, $k$ the potential
curvature and $a$ the particle radius. Obviously, \taueq is entirely
determined by the state into which the relaxation occurs and
independent of the initial conditions. In order to investigate
whether this holds also for relaxation processes into nonequilibrium
states, we systematically vary the initial \NESSII (by changing
$\fII$ and $\phi_{\RM{2}}$) and study the relaxation into the
identical final \NESSI (same parameters as in Fig.~\ref{fig:histo}).
As an example, Fig.~\ref{fig:zerfall}(c) shows the relaxation for
$\fII\approx 52\unit{\kT/\mu m}$ and $\phi_{\RM{2}}=\pi$. It should
be realized that although the steady state distribution of the
initial state in Figs.~\ref{fig:zerfall}(b) and (d) is rather
different, the decay of the mean drift velocity is - within our
experimental errors being caused by the finite number of
trajectories and small optical drifts - identical. This is also seen
in Table~\ref{tab:zerfallszeit} which summarizes five relaxation
experiments from different \NESSII into the identical \NESSI. Within
our experimental accuracy we observe the same relaxation time
$\tauR=1.6\pm 0.2\unit{s}$. At least in case of the specific NESS as
considered here, this suggests that the relaxation time only depends
on the final state. We confirmed the independence of the relaxation
time for a variety of different \NESSI conditions. Due to technical
details all these experiments were performed in the running regime.

\begin{table}

\begin{tabular}{|c|c|c|c|c|}
 \hline
  & $\fII$ & $\phi_{\text{\RM{2}}}$ & $\vII$  & $\tau$ \\
  &   $[\unit{\kT/\mu m}]$ &  &  $[\unit{\mu m / s}]$ & $[\unit{s}]$ \\
 \hline
 (i)& $4$& $\pi$& $0$& $1.4\pm 0.2$\\
 \hline
 (ii)&$4$&$0$&$0$&$1.6\pm 0.25$\\
 \hline
 (iii) & $38$ & $\pi$ & $3.4$ & $1.6\pm 0.25$\\
 \hline
 (iv) & $52$ & $\pi$ & $9.7$ & $1.9 \pm 0.3$\\
 \hline
 (v) & $99$ & $\pi$ & $21$ & $1.6\pm 0.25$ \\
 \hline
\end{tabular}
\caption{Measured relaxation time of NESS~\RM{1} for different
parameters of the initial \NESSII.\label{tab:zerfallszeit}}
\end{table}

\begin{figure}
  \includegraphics{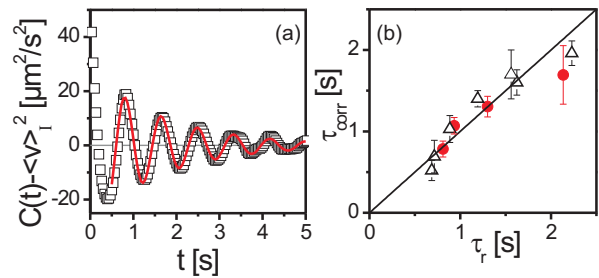}
\caption{(color online) (a) Measured velocity autocorrelation
function (symbols). The solid line is an exponentially decaying
sinusoidal fit. (b) Relaxation time $\tauR$ vs. \tauCorr  for
different NESS where the driving force has been varied between $40$
and $87\unit{\kT/\mu m}$. The open and closed symbols correspond to
potential depths of $125$ and $100\unit{\kT}$, respectively. The
straight line has slope one.\label{fig:autocorr}}
\end{figure}


According to the fluctuation dissipation theorem (FDT), the temporal
decay of fluctuations does not depend on whether they are imposed by
an external force or spontaneously generated by the system itself.
However, it is important to realize that this identity is only valid
in or close to thermal equilibrium~\cite{ris96}. Therefore it is not
{\it a priori} clear whether in a driven system (as considered here)
\tauR, i.e. the decay time in response to a sudden change of the
external driving force is identical with the decay of the steady
state fluctuations. As shown in Fig.~\ref{fig:autocorr}(a) the mean
drift velocity autocorrelation function
$C(t)=\mean{v(t')v(t'+t)}_{t'}$, obtained via a stationary
measurement under \NESSI conditions, is an exponentially decaying
sinusoidal function with a decay time of $\tauCorr=1.7\pm
0.2\unit{s}$. Within the experimental error this value is again
identical with the above determined \tauR. To test whether this
agreement is generally valid, we performed additional measurements
with different driving forces $f$ and potential depths $V_0$. In
Fig.~\ref{fig:autocorr}(b) we compare $\tauCorr$ and $\tauR$ as
measured for $40\unit{\kT/\mu m}<f_{I}<87\unit{\kT/\mu m}$ and
$V_0=100\unit{\kT}$ and $V_0=125\unit{\kT}$, respectively. The good
agreement between the data points and the solid line (slope one)
supports that $\tauR=\tauCorr$ and suggests that (as in
equilibrium), the relaxation time of a NESS can be measured via
transient or stationary measurements.


In order to compare the relaxation time with theory, we calculate
$\tauR$ by numerically solving the Fokker-Planck equation
~\cite{ris96}
\begin{equation}
\partial_{t}\,p(x,t)=-\partial_x[\mu_0
F(x)-D_0\partial_x]\,p(x,t),\label{eq:fpneu}
\end{equation}
with $R$ the torus radius and $F(x)=-\frac{\partial U(x)}{\partial
x}$ the total external force acting on the particle. The transport
coefficients are assumed to be uneffected by the external driving
force~\cite{sei08, spe06}, therefore the free diffusion coefficient
$D_0$ and the mobility $\mu_0$ are taken from thermal equilibrium.
In units of dimensionless time $\tilde{t}=(D_0/R^2)t\equiv
\varepsilon^{-1}t$ and force $\tilde{F}(x)=(R/\kT)F(x)$ the
Fokker-Planck equation reduces to $\partial _{\tilde{t}}\,
p(x,\tilde{t})=\hat{L}_{x} p(x ,\tilde{t})$. Since the Fokker-Planck
operator $\hat{L}_{x}=-\partial_x\tilde{F}(x)+\partial^2_{x}$ has no
explicit time dependence, a separation ansatz for the probability
distribution $p(x ,\tilde{t})=\sum_n \exp(-\lambda_n \,\tilde{t})
q_n(x)$ leads to the following eigenvalue equation
\begin{equation}
-\lambda_n\,q_n(x)= \hat{L}_{x}\,q_n(x).\label{eq:separation}
\end{equation}

The relaxation of an arbitrary given initial probability
distribution is described by the complete set of eigenvalues
$\lambda_n$. However, in the long time limit only the two smallest
eigenvalues $\lambda_0$ and $\lambda_1$ are relevant. The stationary
solution $p^{\text{s}}(x)$ is given by $\lambda_0=0$ and $q_0(x)$.
The real part of $\lambda_1$,
$\mathcal{R}(\lambda_1)\equiv\tau_1^{-1}$ determines the asymptotic
time dependence of the relaxation process. Therefore the relaxation
time is $\varepsilon\tau_1$. Since Eq.~\eqref{eq:separation} has no
analytical solution, for the determination of the eigenvalues we
have to expand the eigenfunctions into an orthonormal basis. Due to
the periodic nature of the system a suitable choice is the Fourier
series $q_n(x)=\frac{1}{\sqrt{2\pi}}\sum_l c_l^{(n)} \exp(ilx)$. A
straightforward calculation leads to
\begin{equation}
-\lambda_n c_k^{(n)}=\sum_l L_{kl}c_l^{(n)},
\end{equation}
an eigenvalue equation for the matrix $\textbf{L}\equiv(L_{kl})$. In
case of the experimentally realized sinusoidal potential, $\bf{L}$
is tridiagonal \cite{mehl08}. After truncating the size of the
matrix to a finite value its eigenvalues are easily found using
standard numerical algorithms.

\begin{figure}
  \includegraphics[width=6.6cm]{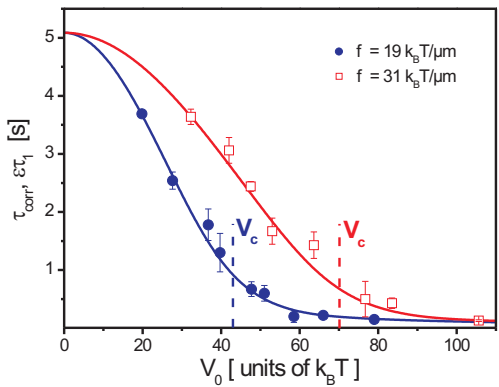}
\caption{(color online) Symbols: Measured decay times \tauCorr for
two different driving forces. The solid lines show the
parameter-free numerical prediction for $\varepsilon\tau_1$. For
potential depths smaller than the critical amplitude
$V_{\text{c}}=2fR$ the minimum in the tilted potential vanishes,
which leads to an enhancement of the diffusion
coefficient~\cite{rei01}. \label{fig:relaxzeit}}
\end{figure}

Fig.~\ref{fig:relaxzeit} shows the calculated $\varepsilon\tau_1$
(solid lines) as a function of the potential depth $V_0$ and for two
different driving forces $f$. For large $V_0$ the system becomes
equilibrium like and the relaxation time asymptotically approaches
the Ornstein-Uhlenbeck result of Eq.~\eqref{eq:tauRgg}. At small
$V_0$ the relaxation time is only determined by the timescale the
particle needs to diffuse along the toroidal trap, i.e.
$\varepsilon\tau_1=D_0/R^2$. These two limiting cases are connected
via a monotonic curve. The closed symbols correspond to the
experimentally determined decay time \tauCorr. The excellent
parameter free agreement between experimental and numerical results
again supports our assumption that \tauCorr is equal to \tauR (see
Fig.~\ref{fig:autocorr}) and {\it a posteriori} justifies that even
for driven colloidal systems the Fokker-Planck equation, with $D_0$
taken from equilibrium, is still valid.

In summary, we have investigated the relaxation behavior of a
colloidal particle into a NESS. Our results show that the NESS
relaxation time is independent of the initial conditions from
which the relaxation process starts. In agreement with
calculations we confirm that, in case of a driven colloidal
particle, the nonequilibrium relaxation time is identical to the
decay time of the velocity autocorrelation function. It must be
emphasized that it is not clear yet whether our observations are
generally valid to arbitrary NESS or restricted to particular
situations. We hope that our work will stimulate further
theoretical studies in this direction. It will be also interesting
to perform similar stationary nonequilibrium relaxation
measurements in systems of sheared polymers~\cite{gera08} or
vesicles~\cite{kan08}.

We thank Udo Seifert and Thomas Speck for fruitful discussions and
suggestions. V.B. was supported by the Deutsche
Forschungsgemeinschaft (BL-1067).


\end{document}